\documentclass{IEEEtran}
\usepackage{lineno}
\usepackage{amssymb}
\usepackage{amsfonts}
\usepackage{amsmath}
\usepackage{algorithm}
\usepackage{algorithmic}
\usepackage{tikz}
\usepackage{graphicx,subfigure,cite,multicol,multirow,diagbox,booktabs,array}
\usepackage{color}
\usepackage{bm}
\usetikzlibrary{shapes.geometric, arrows}
\ifCLASSINFOpdf
\else
\fi

\begin{document}
\bibliographystyle{IEEEtran}

\title{High-performance Estimation of Jamming Covariance Matrix for IRS-aided Directional Modulation Network with a Malicious Attacker }

\author{Hangjia He,~Ting Su,~Hongjun~Wang,~Yin Teng,~Weiping Shi,~Feng Shu,\\
and Jiangzhou~Wang,~\IEEEmembership{Fellow,~IEEE}
\thanks{Hangjia He,~Yin Teng,~and Weiping Shi are with the School of Electronic and Optical Engineering, Nanjing University of Science and Technology, Nanjing, 210094, China. }
\thanks{Ting Su and Feng Shu are with the School of Information and Communication Engineering, Hainan University, Haikou 570228, China. and also with the School of Electronic and Optical Engineering, Nanjing University of Science and Technology, Nanjing 210094, China. (e-mail: shufeng0101@163.com;suting4190@hainanu.edu.cn). }
\thanks{Hongjun Wang is with the School of Information and Communication Engineering,  National University of Defense Technology, China.  (e-mail: hongjun-wang@163.com). }
\thanks{Jiangzhou Wang is with the School of Engineering and Digital Arts, University of Kent, Canterbury CT2 7NT, U.K. (e-mail: j.z.wang@kent.ac.uk).}}
\maketitle

\begin{abstract}
In this paper, we investigate the anti-jamming problem of a directional modulation (DM) system with the aid of intelligent reflecting surface (IRS). As an efficient tool to combat malicious jamming, receive beamforming (RBF) is usually designed to be on null-space of jamming channel or covariance matrix from Mallory to Bob. Thus, it is very necessary to estimate the receive jamming covariance matrix (JCM) at Bob. To achieve a precise JCM estimate, three JCM estimation methods, including eigenvalue decomposition (EVD), parametric estimation method by gradient descend (PEM-GD) and parametric estimation method by alternating optimization (PEM-AO), are proposed. Here, the proposed EVD is  under rank-2 constraint of JCM. The PEM-GD method fully explores the structure features of JCM and the PEM-AO is to decrease the computational complexity of the former via dimensionality reduction. The simulation results show that in low and medium jamming-noise ratio (JNR) regions, the proposed three methods perform better than the existing sample covariance matrix method. The proposed PEM-GD and PEM-AO outperform EVD method and existing clutter and disturbance covariance estimator RCML.

\end{abstract}


\section{Introduction}
In wireless communication systems, due to the borderless feature of radio propagation, confidential messages (CM) may be tapped by eavesdroppers, thus research on physical layer security (PLS) is essential for secure transmission. Among multiple techniques of PLS, directional modulation (DM) is a valid scheme which can send signals directionally and purposely distort the signals in other directions{\cite{dm2009,FDADM,robustDM,Precise,multibeam}}. Furthermore, while wireless communication develops rapidly in recent years, high hardware complexity as well as energy consumption is a critical issue yet{\cite{wu2020}}. Under such circumstance, the intelligent reflecting surface (IRS){\cite{wu2020,DMlai,IRSDM,chengRIS}} is believed to be a promising new technology, which can smartly reconfigure the wireless propagation environment at lower cost.
When combining DM and IRS, the communication system can achieve higher performance than conventional DM {\cite{DMlai}}. And since IRS can create friendly multipaths for DM, it is possible to transmit two or more confidential bit streams with the aid of IRS in DM, the authors of {\cite{IRSDM}} proposed this scheme, and it may greatly increase the secrecy rate (SR) of transmission.

While the aforementioned works all focused on preventing CM leakage, the receiver may also be subject to malicious jamming. In {\cite{yang2020intelligent}}, the authors proposed a learning approach to resist jamming by jointly optimizing the transmit power allocation and the reflecting beamforming matrix in an IRS assisted system. This work deals with the situation where receivers are equipped with single antenna, and when receivers are equipped with multiple antennas, receive beamforming (RBF) is an efficient anti-jamming scheme. The authors in {\cite{teng2020low}} and {\cite{jiang2021efficient}} presented scenarios with a full-duplex (FD) malicious attacker Mallory, and they proposed several RBF methods, which can solve the anti-jamming problem with high-performance.

In this paper, we consider an IRS-aided DM network with a malicious attacker where Alice, Mallory, and Bob are equipped with multiple antennas. RBF methods in {\cite{teng2020low}} can be applied to eliminate jamming from Mallory. However, how to estimate the channel state information (CSI) of jamming channel or the statistical property of jamming signal is the key to design RBF at Bob. Here, Mallory is a non-cooperative unit, so we need to estimate  the jamming covariance matrix (JCM) from Mallory at Bob. Thus, three methods are proposed to estimate the JCM from sample covariance matrix (SCM). Our main contributions are summarized as follows

\begin{enumerate}
    \item To estimate JCM precisely, minimizing the Euclidean distance between estimated JCM and sample covariance matrix under different constraints is established as an optimization rule. The rank of ideal JCM is derived to be two, and an eigenvalue decomposition (EVD) method is proposed with rank-2 as a rank constraint. Simulation results show that the proposed EVD method performs better than existing method of directly using the definition of SCM, called SCM, but it is inferior to RCML in {\cite{kang2014rank}}. However, RCML requires the knowledge of receive noise variance while the proposed EVD may estimate receiver noise variance. Thus, the proposed EVD is more practical.
    \item To achieve a  better estimation, we then exploit the structure properties of JCM. By observing the expression of ideal JCM, we extract the unknown parameters, integrate and decompose them to four vectors, and then the estimation problem is converted into a problem of optimizing four unknown vectors, which forms a parametric estimation method by gradient descend (PEM-GD). The JCM estimated by the proposed PEM-GD is independent of the phase changes of the IRS. To reduce the complexity of PEM-GD, a  dimensionality-reduction method with fewer optimization variables, called PEM by alternating optimization (PEM-AO), is proposed. Simulation results show that the proposed PEM-GD and PEM-AO have the same NMSE performance with the latter being lower-complexity, and outperform EVD and RCML.
\end{enumerate}

The remainder is organized as follows. Section \ref{model} presents the system model and three estimation methods are proposed in Section \ref{method}. In Section \ref{simulation}, numerical simulations are presented, and Section \ref{conclusion} draws our conclusion.

\emph{Notations:} In this paper, matrices, vectors, and scalars are denoted by uppercase bold, lowercase bold, and lowercase letters, respectively. Signs $(\cdot)^H$, $(\cdot)^T$ $tr(\cdot)$ and $\rm E[\cdot]$ stand for the conjugate transpose, transpose, trace and expectation operation respectively. $\|\cdot\|_F$ denotes the Frobenius norm of a matrix, and $\Re \{\cdot\}$ represents the real part of a variable.

\section{System Model}\label{model}

\begin{figure}[ht]
\centering
\includegraphics[scale=0.5]{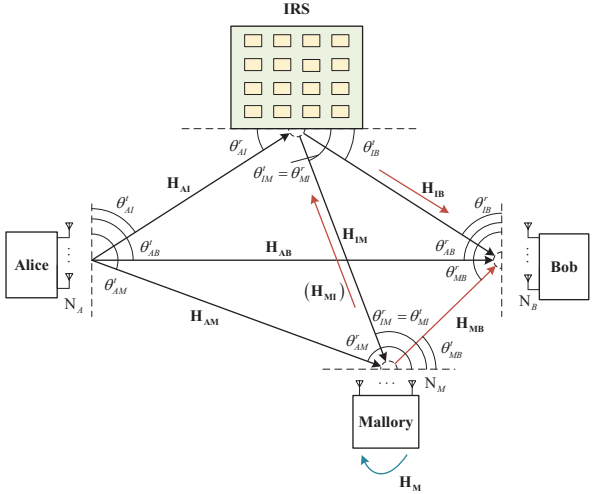}
\centering
\caption{Block diagram of IRS-aided DM network with malicious attacker}
\end{figure}

Fig.1 shows an IRS-aided DM wireless communication, where a transmitter (Alice) equipped with $N_A$ antennas sends CM to a legitimate user (Bob) with $N_B$ antennas. The transmission is assisted by IRS with $M$ passive reflecting elements. It is assumed that the phase shift of the IRS and transmit beamforming at Alice have been optimized and fixed when an illegal malicious attacker Mallory is detected. Mallory with $N_M$ antennas works at FD mode, which means that it can eavesdrop the message from Alice and send malicious jamming to Bob simultaneously.

The transmitted baseband signal from Alice is
\begin{align}
    {\bf s}_A = \sqrt {\beta {P_A}} {\bf{v}}x + \sqrt {(1 - \beta ){P_A}} {\bf T}_{A,AN}{\bf z}_{A,AN},
\end{align}
where $P_A$ denotes the total transmit power and $\beta \in [0, 1]$ is the power allocation factor. ${\bf v} \in \mathbb{C}^{N_A\times 1}$ and ${\bf T}_{A,AN} \in \mathbb{C}^{N_A\times N_A}$ denote the transmit beamforming vector and projection of AN respectively, with the nature of ${\bf v}^H {\bf v} = 1$ and $tr({\bf T}_{A,AN}{\bf T}_{A,AN}^H)=1$, and AN is designed in accordance with the null-space projection. $x$ is the transmitted symbol satisfying ${\rm E}[|x|^2]=1$, and ${\bf z}_{A,AN} \in \mathbb{C}^{N_A \times 1}$ represents the AN vector with complex Gaussian distribution, ${\bf z}_{A,AN} \sim {\mathcal {CN}}({\bf0, I}_{N_A})$.

The malicious jamming signal from Mallory is
\begin{align}
    {\bf s}_M = \sqrt {{P_M}} {\bf T}_{M,AN}{\bf z}_{M,AN},
\end{align}
where $P_M$ is the transmit power of Mallory , ${\bf T}_{M,AN} \in \mathbb{C}^{N_M\times N_J}$ denotes the projection of jamming, $N_J \in [1, N_M-1]$, and
${\bf z}_{M,AN} \sim {\mathcal {CN}} {({\bf 0,I}_{N_J})}$ indicates the jamming symbol from Mallory with complex Gaussian distribution.

The received signal at Bob can be written as:
\begin{align}\label{eq0rb}
    {\bf r}_B &= {\bf v}_{BR}^H [(\underbrace{{\sqrt {g_{AIB}}
        {\bf H}_{IB}^H}{{\boldsymbol \Theta}}{\bf H}_{AI}^H
        + \sqrt {{g_{AB}}} {\bf H}_{AB}^H}_{{\bf H}_{A1}}){\bf s}_A\nonumber\\
    &+ (\underbrace{{\sqrt {g_{MIB}} {\bf H}_{IB}^H}
        {\boldsymbol \Theta}{\bf H}_{MI}^H
        + \sqrt {g_{MB}} {\bf H}_{MB}^H}_{{\bf H}_{M1}}){\bf s}_M + {\bf n}_B ]\nonumber\\
    &= {{\bf v}_{BR}^H}[\sqrt {\beta {P_A}}{{\bf H}_{A1}}{\bf{v}}x
    + \underbrace {{\sqrt {(1 - \beta ){P_A}}
        {{\bf H}_{A1}}}{{\bf T}_{A,AN}} {{\bf z}_{A,AN}}}_{{\bf n}_{AB}}\nonumber\\
    &+ \underbrace {{\sqrt {{P_M}} {{\bf H}_{M1}}}
        {{\bf T}_{M,AN}}{{\bf z}_{M,AN}}}_{{\bf n}_{MB}}
        + {\bf n}_B],
\end{align}
where ${\bf v}_{BR} \in \mathbb{C}^{N_B \times 1}$ is the receiving beamforming vector of Bob, ${\bf n}_B \in \mathbb{C}^{N_B \times 1}$ denotes the complex additive white Gaussian noise (AWGN) vectors, following the distribution ${\bf n}_B \sim {\mathcal {CN}}({\bf 0},\sigma_B^2 {\bf I}_{N_B})$, and ${\boldsymbol \Theta} = diag\{[e^{j\phi_1},..., e^{j\phi_i},..., e^{j\phi_M}]\}$ is a diagonal matrix with $\phi_i$ symbolizing the phase shift of the ith element at the IRS. $g_{AIB}$, $g_{AB}$, $g_{MIB}$, and $g_{MB}$ denote the path loss coefficients of four path: Alice to Bob through IRS, Alice to Bob directly, Mallory to Bob through IRS and Mallory to Bob directly.

Besides, ${\bf H}_{IB}^H \in \mathbb{C}^{N_B \times M}$, ${\bf H}_{AI}^H \in \mathbb{C}^{M \times N_A}$, ${\bf H}_{AB}^H \in \mathbb{C}^{N_B \times N_A}$, ${\bf H}_{MI}^H \in \mathbb{C}^{M \times N_M}$, and ${\bf H}_{MB}^H \in \mathbb{C}^{N_B \times N_M}$ denote the channel matrices from IRS to Bob, Alice to IRS, Alice to Bob, Mallory to IRS and Mallory to Bob respectively. Since in DM network, transmitter and receiver are deployed with $N$-element linear antenna arrays, the normalized steering vector is given by
\begin{align}
    {\bf{h}}(\theta ) = \frac{1}{{\sqrt N }}
        {\left[ {{e^{j2\pi {\Psi _\theta }(1)}},...,{e^{j2\pi {\Psi _\theta }(n)}},...,{e^{j2\pi {\Psi _\theta }(N)}}} \right]^{T}},
\end{align}
where
\begin{align}\label{eq5}
    {\Psi _\theta }(n) =  - (n - \frac{{N + 1}}{2})\frac{{d\cos \theta }}{\lambda }\qquad n = 1...N.
\end{align}
where $\theta$ represents the angle of arrival or departure of signal, n is the index of antenna, while $N$ is the number of antennas. $d$ is the separation distance of antenna array and $\lambda$ represents the wavelength. Then, the channel can be given by ${\bf H}^H (\theta) = {\bf h}(\theta_r){\bf h}^H (\theta_t)$. In (\ref{eq0rb}), for convenience, we set ${\bf H}_{A1}$ as the equivalent channel matrix of Alice to Bob, and ${{\bf H}_{M1}}$ as the equivalent channel matrix of Mallory to Bob.

The receive JCM from Mallory at Bob is
\begin{align}
\label{ri}
    {\bf R}_{i} &= {\rm E}\left [ {\bf n}_{MB} {\bf n}_{MB}^H\right ]\nonumber\\
    &= {P_M}(\sqrt {{g_{MIB}}} {\bf H}_{IB}^H{\boldsymbol \Theta }{\bf H}_{MI}^H
    + \sqrt {{g_{MB}}} {\bf H}_{MB}^H){\bf T}_{M,AN} \cdot\nonumber\\
    &
    {\bf T}_{M,AN}^H(\sqrt {{g_{MIB}}} {\bf H}_{IB}^H {\boldsymbol \Theta }{\bf H}_{MI}^H
        + \sqrt {g_{MB}} {\bf H}_{MB}^H)^H.
\end{align}
To estimate  JCM,  once Alice detects the jamming signal from Mallory, she will keep silent, and then the received jamming signal plus noise at Bob is as follows
\begin{align}
{\bf y}_B
    = \underbrace {{\sqrt {{P_M}} {\bf H}_{M1}}
       {\bf T}_{M,AN}{\bf z}_{M,AN}}_{{\bf n}_{MB}}
        + {\bf n}_B.
\end{align}
After Bob receives $K$ samples, SCM can be directly given  by
\begin{align}\label{SamCM-def}
    {\bf{\hat R}} &= \frac{1}{K}\sum\limits_{k = 1}^K {\bf y}_B [k]{\bf y}_B^H [k].
\end{align}
As $K$ tends to infinity,
\begin{align}
    {\bf{\hat R}}\approx {\bf R}_{i}+\sigma_B^2 {\bf I}_{N_B},
\end{align}
which means, when $K$ is large enough, (\ref{SamCM-def}) is a valid estimator. By exploring the property of statistical covariance matrix (\ref{ri}), we find it has some excellent properties like rank-2 or rank-1 channel matrix. This will be utilized to improve the poor performance of SCM method in (\ref{SamCM-def}) in the small-sample scenario. In the following, the criterion of minimizing the Euclidean distance between SCM minus noise covariance matrix ${\bf S} = \hat{\bf R} -\sigma_B^2 {\bf I}_{N_B}$ and estimated JCM is cast as
\begin{align}\label{org-opt}
    &\mathop{\min}\limits_{\bf R} \quad \| {\bf R} - {\bf S} \|_F,
\end{align}
subject to the constraints of JCM properties.

To compare the SR performance of JCM estimated by different schemes, we extend the NSP-based Max-WFRP RBF  method in {\cite{teng2020low}} to our model, which can be cast as
\begin{align}
\begin{split}
    &\max_{{\bf v}_{BR}} \quad
        {\bf v}_{BR}^H {\bf H}_{A1} {\bf v} {\bf v}^H {\bf H}_{A1}^H {\bf v}_{BR} \\
    &\text{s.t.} \quad {\bf v}_{BR}^H {\bf R} = {\bf 0}_{1 \times N_B},
         {\bf v}_{BR}^H {\bf v}_{BR} =1.
\end{split}
\end{align}

\section{Proposed Three Estimation Methods}\label{method}
In this section, by exploring the features of ideal JCM, including its rank and decomposition, three estimate methods named EVD, PEM-GD and PEM-AO are proposed to improve the NMSE performance, and their complexities are also compared.
\subsection{Proposed EVD method}
 Let us first consider the rank-2 constraint of the JCM ${\bf R}_{i}$ in (\ref{ri}), it can be expanded as
\begin{align}
    {\bf R}_{i} = \lambda_{r1}{\bf v}_{r1}{\bf v}_{r1}^H
        +\lambda_{r2}{\bf v}_{r2}{\bf v}_{r2}^H \label{r2},
\end{align}
where $\lambda_{r1},\lambda_{r2}$ are the eigenvalues of ${\bf R}_i$, and ${\bf v}_{r1},{\bf v}_{r2}$  the corresponding eigenvectors. Meanwhile, the covariance matrix of ${\bf y}_B$ is given by
\begin{align}
\begin{split}
    {\bf R}_y &= {\rm E}[{\bf y}_B {\bf y}_B^H]={\bf R}_{i} + \sigma_B^2{\bf I}_{N_B}.
\end{split}
\end{align}
whose eigenvalues have the order ${\lambda}_{y1} \ge ... \ge {\lambda}_{yN_B}$,  and ${\bf v}_1, ..., {\bf v}_{N_B}$ are the corresponding eigenvectors of ${\bf R}_y$. Here, ${\lambda}_{y1} = \lambda_{r1} + {\sigma_B^2}, {\lambda}_{y2} = \lambda_{r2} + {\sigma_B^2}$, and ${\lambda_{y3}} = ... = {\lambda_{yN_B}} = {\sigma_B^2}$ and ${\bf v}_1 ={\bf v}_{r1}$, ${\bf v}_2 ={\bf v}_{r2}$, i.e. ${\bf R}_y$ can be rewritten as
\begin{align}
\label{ry}
    {\bf R}_y &= (\lambda_{r1} + \sigma_B^2){\bf v}_{1}{\bf v}_{1}^H
        +(\lambda_{r2} + \sigma_B^2){\bf v}_{2}{\bf v}_{2}^H
        + \sum_{i=3}^{N_B} \sigma_B^2 {\bf v}_i{\bf v}_i^H,
\end{align}

Since ${\bf R}_y$ and ${\bf{\hat R}}$ give the expectation and sample means of ${\bf y}_B$ respectively, after finding the eigenvalues and eigenvectors of ${\bf{\hat R}}$, represented as ${\lambda}_1 \ge ... \ge {\lambda}_{N_B}$ and ${\bf u}_1, ..., {\bf u}_{N_B}$, the  receive noise variance and JCM are estimated as
\begin{align}
    \hat {\sigma}_B^2 = \frac{ \sum_{i=3}^{N_B} \lambda_i}{N_B - 2},
\end{align}
and
\begin{align}
    {\bf R}_{EVD} =
        (\lambda_1 - \hat {\sigma}_B^2){\bf u}_1{\bf u}_1^H
        +(\lambda_2 - \hat {\sigma}_B^2){\bf u}_2{\bf u}_2^H.
\end{align}

\subsection{Proposed PEM-GD}
Now, we turn to consider the structure of ${\bf R}_{i}$ and propose a method to estimate JCM by its parameters. It can be derived from (\ref{ri}) that ${\bf R}_{i} = {\bf FF}^H$, where
\begin{align}
    {\bf F} &= \sqrt {P_M} (\sqrt {{g_{MIB}}} {\bf H}_{IB}^H{\boldsymbol \Theta }
        {\bf H}_{MI}^H+ \sqrt {{g_{MB}}}{\bf H}_{MB}^H){\bf T}_{M,AN}\nonumber\\
    &= {\bf H}_{IB}^H{\boldsymbol \Theta} \underbrace {(\sqrt{P_M g_{MIB}}{\bf H}_{MI}^H {\bf T}_{M,AN}) }_{{\bf T}_1}\nonumber\\
    &+ \underbrace {\sqrt{P_M g_{MB}}{\bf H}_{MB}^H {\bf T}_{M,AN}}_{{\bf T}_2}.
\end{align}
where matrices ${\bf T}_1 \in \mathbb{C}^{M \times N_J}$ and ${\bf T}_2 \in \mathbb{C}^{N_B \times N_J}$ describe the unknown parameters in ${\bf R}_{i}$. Since both ${\bf T}_1$ and ${\bf T}_2$ are rank-one matrices, they can be decomposed into ${\bf T}_1 = {\boldsymbol {\alpha \beta}}^H, {\bf T}_2 = {\boldsymbol {\omega \nu}}^H$, where ${\boldsymbol \alpha} \in \mathbb{C}^{M \times 1}, {\boldsymbol \omega} \in \mathbb{C}^{N_B \times 1}$, and $\boldsymbol {\beta,\nu} \in \mathbb{C}^{N_J \times 1}$. Therefore, the estimated JCM  is constructed as
\begin{align}
\label{rabwv}
\begin{split}
    {\bf R}(\boldsymbol {\alpha, \beta, \omega, \nu})&=
        ({\bf H}_{IB}^H {\boldsymbol {\Theta \alpha \beta}}^H + {\boldsymbol \omega \nu}^H) \cdot \\
    &\quad ({\bf H}_{IB}^H {\boldsymbol {\Theta \alpha \beta}}^H + {\boldsymbol \omega \nu}^H)^H.
\end{split}
\end{align}
which transforms the optimization problem in (\ref{org-opt}) into
\begin{align}
\label{f}
    \mathop{\min}\limits_{\boldsymbol {\alpha, \beta, \omega, \nu}} \quad
        \| {\bf R(\boldsymbol {\alpha, \beta, \omega, \nu})}- {\bf S} \|_F^2,
\end{align}
it is  an unconstrained non-convex optimization problem, then the gradient descend method (GD) is applied to get the unknown parameters in (\ref{rabwv}). The gradients of the objective function with respect to $\boldsymbol \alpha$, $\boldsymbol \beta$, $\boldsymbol \omega$ and $\boldsymbol \nu$ are as follows
\begin{align}
\begin{split}
    \nabla_{\boldsymbol \alpha^*} &= {\boldsymbol \Theta}^H {\bf H}_{IB}{\bf (R - S)}^H
        ({\bf H}_{IB}^H {\boldsymbol{\Theta \alpha \beta}}^H + {\boldsymbol{\omega \nu}}^H){\boldsymbol \beta} ,\\
    \nabla_{\boldsymbol \beta^*} &= ({\boldsymbol{\nu \omega}}^H
        + {\boldsymbol {\beta \alpha}}^H {\boldsymbol \Theta}^H {\bf H}_{IB}){\bf (R - S)}^H {\bf H}_{IB}^H {\boldsymbol{\Theta \alpha}} ,\\
    \nabla_{\boldsymbol \omega^*} &= {\bf (R-S)}^H({\bf H}_{IB}^H {\boldsymbol
        {\Theta \alpha \beta}}^H + {\boldsymbol{\omega \nu}}^H){\boldsymbol \nu }, \\
    \nabla_{\boldsymbol \nu^*} &= ({\boldsymbol {\beta \alpha}}^H
        {\boldsymbol\Theta }{\bf H}_{IB}+{\boldsymbol {\nu \omega}}^H){\bf(R-S)}^H {\boldsymbol \omega}.
\end{split}
\end{align}

In the above GD algorithm, we first initialize the parameters $\boldsymbol \alpha$, $\boldsymbol \beta$, $\boldsymbol \omega$ and $\boldsymbol \nu$, and noting that $N_J$ is unknown in a practical scene, $\boldsymbol \beta$ and $\nu$  are initialized to vectors of dimension $\tilde{N}_J$  greater than the surmised number of Mallory's antennas. Next,  all parameters are updated as ${\bf x}^{(m)} = {\bf x}^{(m-1)} + t_{\bf x}^{(m)} \nabla_{\bf x*}^{(m-1)} $ in each iteration until convergence, where $\bf x$ can be replaced by $\boldsymbol \alpha$, $\boldsymbol \beta$, $\boldsymbol \omega$ and $\boldsymbol \nu$. It
should be aware that $t_{\bf x}$ denotes the step of each update, which can be obtained by a backtracking line search in {\cite{boyd2004convex}}, and it guarantees that the objective function declines in each iteration. Thus, the estimated JCM ${\bf R}_{PEM-GD}$ can be obtained, and it can adapt to the phases change of IRS since the estimated vectors are independent of $\boldsymbol \Theta$.

\subsection{Proposed PEM-AO method}
However, while backtracking line search and a sufficiently large initial step size in each search make it possible to get a global minimum point in the previous PEM-GD method, the GD method of four vectors and backtracking line search cause a large computational amount. Therefore, below, improving the work in the previous subsection, a lower complexity parametric estimation method is proposed. From ({\ref{rabwv}}), the estimated JCM is given by
\begin{align}
    &\quad {\bf R}(\boldsymbol {\alpha, \beta, \omega, \nu})
    ={\bf H}_{IB}^H {{\boldsymbol {\Theta \alpha \beta}}^H {\boldsymbol {\beta \alpha}}^H {\boldsymbol \Theta}^H} {\bf H}_{IB}\\
    &+ {\bf H}_{IB}^H {{\boldsymbol {\Theta \alpha \beta}}^H { \boldsymbol {\nu \omega}}^H}
    + {{\boldsymbol {\omega \nu}}^H {\boldsymbol {\beta \alpha}}^H {\boldsymbol \Theta}^H} {\bf H}_{IB}
    + {{\boldsymbol {\omega \nu}}^H {\boldsymbol {\nu \omega}}^H}.\nonumber
\end{align}

To reduce unknown optimization variables , let us define three new variables ${\boldsymbol\beta}^H {\boldsymbol \beta} =c_1$, ${\boldsymbol \nu}^H {\boldsymbol \beta} = c_2$, ${\boldsymbol  \nu }^H{\boldsymbol \nu} = c_3$, and the association of three newly defined variables can be derived as $c_1 c_3 cos^2 \theta = c_2 c_2^*$, where $\theta$ is the included angle between $\boldsymbol \beta$ and $\boldsymbol \nu$. Since ${\bf H}_{IB}^H = {\bf h}({\theta_{IB}^r}) {\bf h}^H ({\theta_{IB}^t})$, we can set ${\boldsymbol \alpha}^H {\boldsymbol \Theta}^H {\bf h} ({\theta_{IB}^t}) = b$, then the estimated JCM turns into
\begin{align}
\begin{split}
    {\bf R}(c_1,c_2,c_3,b, \boldsymbol {\omega}) =
         c_1 b^* b{\bf h}({\theta_{IB}^r}) {\bf h}^H({\theta_{IB}^r})\\
    + c_2^* b^* {\bf h}({\theta_{IB}^r}) {\boldsymbol \omega}^H
    + c_2  b  {\boldsymbol\omega } {\bf h}^H({\theta_{IB}^r})
    + c_3 {\boldsymbol {\omega \omega}}^H.
\end{split}
\end{align}
And to further reduce the unknown variables, let us define $\tilde{\boldsymbol \omega} = \sqrt{c_3}\boldsymbol \omega$, $\tilde{c}_1 = c_1 b^* b$, $\tilde{c}_2 = \frac{c_2 b}{\sqrt{c_3}}$. Thus, the estimated JCM is formed as
\begin{align}
\begin{split}
    {\bf R}(\tilde{c}_1, \tilde{c}_2, \tilde{\boldsymbol \omega}) &=
        \tilde{c}_1 {\bf h}({\theta_{IB}^r}) {\bf h}^H({\theta_{IB}^r})
        + \tilde{c}_2^* {\bf h}({\theta_{IB}^r}) {\tilde{\boldsymbol \omega}}^H\\
    &+ \tilde{c}_2 \tilde{ \boldsymbol \omega } {\bf h}^H({\theta_{IB}^r})
        + \tilde{ \boldsymbol \omega } {\tilde{ \boldsymbol \omega }}^H,
\end{split}
\end{align}
where $\tilde{c}_1 cos^2 \theta = \tilde{c}_2 \tilde{c}_2^*$.
Consequently, the optimization problem is recast as
\begin{align}
\label{rcw}
    \mathop{\min}\limits_{\tilde{c}_1, \tilde{c}_2, \tilde{\boldsymbol \omega}} ~
        \| {\bf R}(\tilde{c}_1, \tilde{c}_2, \tilde{\boldsymbol \omega}) - {\bf S} \|_{\bf F}^2
    ~ ~ ~ ~\text{s.t.}~ \ \tilde{c}_1 \geq \tilde{c}_2 \tilde{c}_2^*.
\end{align}
Noting that it is hard to solve (\ref{rcw}) directly due to the coupled variables and its non-convex properties, we apply AO algorithm and optimize $(\tilde{c}_1, \tilde{c}_2)$, $\tilde{\boldsymbol \omega}$ alternately. First, by fixing $\tilde{\boldsymbol\omega}$, the sub-optimization problem of $(\tilde{c}_1, \tilde{c}_2)$ is
cast as
\begin{align}
    \mathop{\min}\limits_{\tilde{c}_1, \tilde{c}_2} ~
       f(\tilde{c}_1, \tilde{c}_2)
    ~~~\text{s.t.}~ \tilde{c}_1 \geq \tilde{c}_2 \tilde{c}_2^*,
\end{align}
where
\begin{align}
\begin{split}
    f(\tilde{c}_1, \tilde{c}_2) &= \Re\{\tilde{c}_1^2 + 2(\tilde{c}_2^*)^2a^2 + 4 \tilde{c}_1\tilde{c}_2^*a + 2\tilde{c}_1aa^* \\
    &+ 2\tilde{c}_2\tilde{c}_2^*e + 4\tilde{c}_2^*ae - 2\tilde{c}_1\tau - 4\tilde{c}_2\gamma   \},
\end{split}
\end{align}
with ${\tilde{\boldsymbol \omega}}^H {\bf h}({\theta_{IB}^r}) = a$, ${\tilde{\boldsymbol \omega}^H \tilde{\boldsymbol \omega}} = e$, ${\bf h}^H({\theta_{IB}^r}) {\bf S} {\bf \tilde{\boldsymbol \omega}} = \gamma$, and ${\bf h}({\theta_{IB}^r}){\bf S}{\bf h}({\theta_{IB}^r}) = \tau$ for brevity.

This sub-optimization problem can be solved by the KKT conditions. By setting $m = aa^*-e-v/2$ where $v$ is the Lagrange multiplier associated with the inequality constraint, the result of each iteration is
\begin{align}
    \tilde{c}_1 &= \tau + \frac{v}{2} -aa^*-\tilde{c}_2^*a - \tilde{c}_2a^*,~~ \tilde{c}_2 &= \frac{(\tau -m)a-\gamma^*} {m},
\end{align}
with $v=0$ for $l_3\geq0$, and $v$ being a positive real root of $v^3 + l_1v^2 +l_2v +l_3=0$ when $l_3<0$, where
\begin{align}
\begin{split}
    l_1 &= 4e + 2\tau-4aa^*,\\
    l_2 &= 4a^2(a^*)^2 - 8aa^*e - 8aa^*\tau+4e^2 + 8e\tau,\\
    l_3 &= 8a^2(a^*)^2\tau -16aa^*e\tau -8aa^*\tau^2 + 8\gamma a\tau ,\\
    &+8\gamma^*a^*\tau +8e^2\tau - 8\gamma \gamma^*.
\end{split}
\end{align}
For given $\tilde{c}_1, \tilde{c}_2$, we have to solve a non-convex unconstrained optimization problem about $\tilde{\boldsymbol \omega}$, for which we apply GD method as before, and the gradient of the objective function with respect to $\tilde{\boldsymbol \omega}$ is as follows,
\begin{align}
    \nabla_{\tilde{\boldsymbol \omega}^*} = ({\bf R-S})(\tilde{c}_2^* {\bf h}({\theta_{MB}^r})+ \tilde{\boldsymbol \omega}).
\end{align}

Finally, by alternately calculating $(\tilde{c}_1, \tilde{c}_2)$ and $\tilde{\boldsymbol \omega}$ until convergence, the estimated JCM is obtained as ${\bf R}_{PEM-AO}$.

\subsection{Computational Complexity Analysis and CRLBs}
Now, we analyse the complexities of the proposed methods. The complexities of EVD, PEM-GD, and PEM-AO are $\mathcal{O}(N_B^3 + 2N_B^2+2N_B)$, $\mathcal{O}( L_1(2M^2 N_B + M N_B^2 + M \tilde{N}_J N_B + 3M \tilde{N}_J + 3\tilde{N}_J N_B)log_2(1/\kappa) )$, and $\mathcal{O}(L_2( L_3( N_B^2 + N_B)log_2(1/\kappa) + 3N_B^2 + 2N_B + 25 )+41 L_4)$ respectively, where $L_1$ and $L_3$ denote the the iterative number of GD in PEM-GD and PEM-AO, with $\log_2(1/\kappa)$ the maximum iterative number of backtracking line search, $L_2$ and $L_4$ are the numbers of alternating iterations and occurrences of $l_3<0$ in PEM-AO. Besides, the complexity of RCML in \cite{kang2014rank} is $\mathcal{O}(N_B^3 + 4N_B^2+2)$. Therefore, the complexities of these methods have an decreasing order as PEM-GD, PEM-AO, RCML, and EVD.

Additionally, the CRLBs of JCM is defined as the sum of the Cramer-Rao Lower Bound (CRLB) of each element in JCM to give a lower bound for NMSE. Due to the length limit of paper pages, the CRLBs is directly given by
\begin{align}
    CRLBs ={\|{\bf R}_{i}\|_F^{-2}}{\sum_{j=1}^{N_B} [{\bf I}^{-1}({\bf R})]_{jj}},
\end{align}
with ${\bf I}(\bf R)$ denoting the Fisher information matrix.

\section{Simulation results and Discussions}\label{simulation}

In this section, the performance of the proposed estimation methods are compared through numerical simulations. Simulation parameters are set as follows: $P_A = P_M = 1 W$, $\beta = 0.9$, $N_A = N_B = N_M = 8$, $M = 16$, $\sigma_B^2 = \sigma_M^2$, $K = 5$, $JNR =$5dB and Alice, IRS, Bob, Mallory are located at (0,0), (50,50), (500,0), (400, -50) respectively. SNR and JNR represent the ratio of received jamming to noise and signal to noise respectively.
\setcounter{figure}{2}
\begin{figure*}
\vspace{-0.3cm}
	\setlength{\abovecaptionskip}{-5pt}
	\setlength{\belowcaptionskip}{-10pt}
	\centering
	\begin{minipage}[ht]{0.33\linewidth}
		\centering
		\includegraphics[width=2.56in]{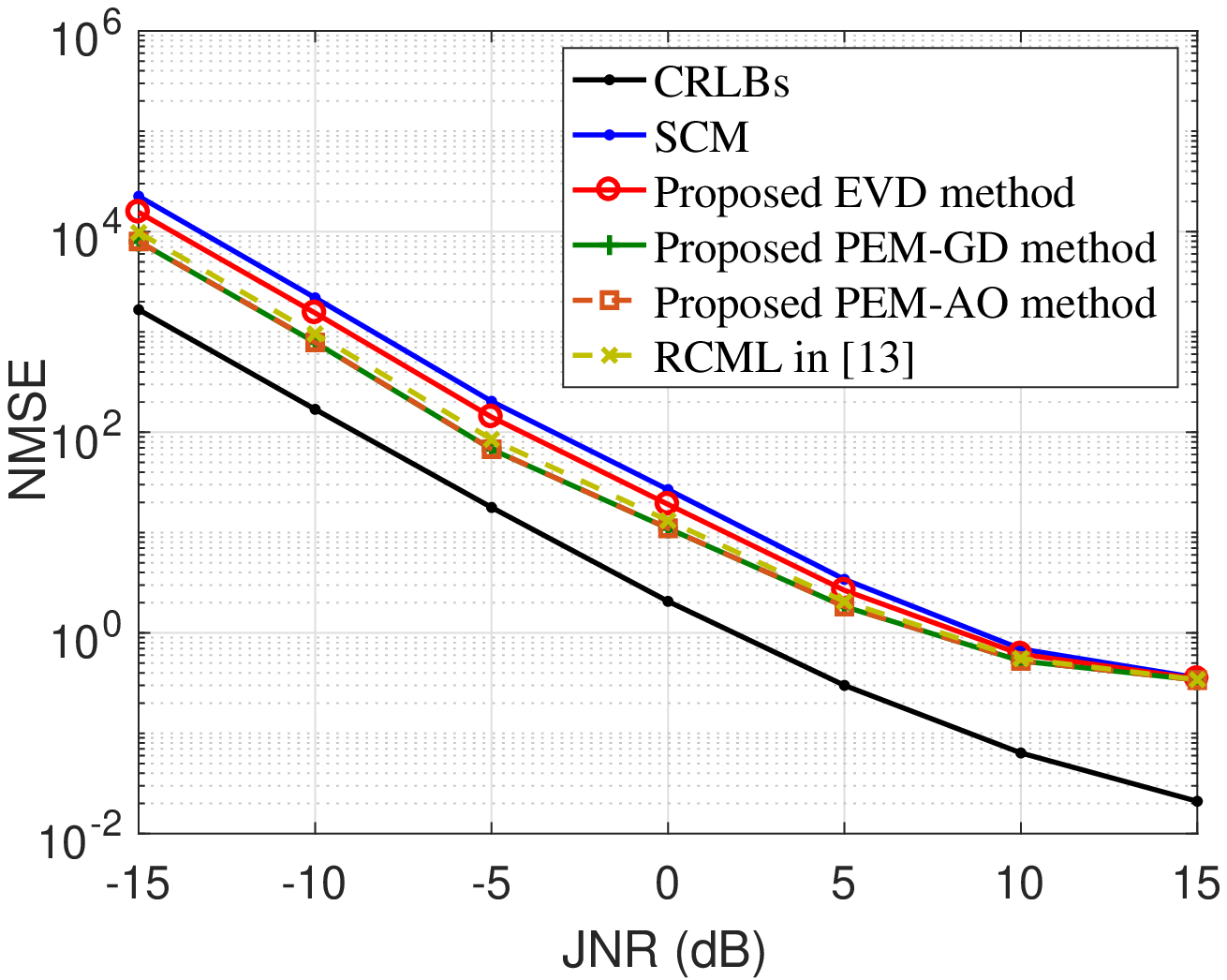}
		\caption{NMSE versus JNR}
	\end{minipage}%
	\begin{minipage}[ht]{0.33\linewidth}
		\centering
		\includegraphics[width=2.56in]{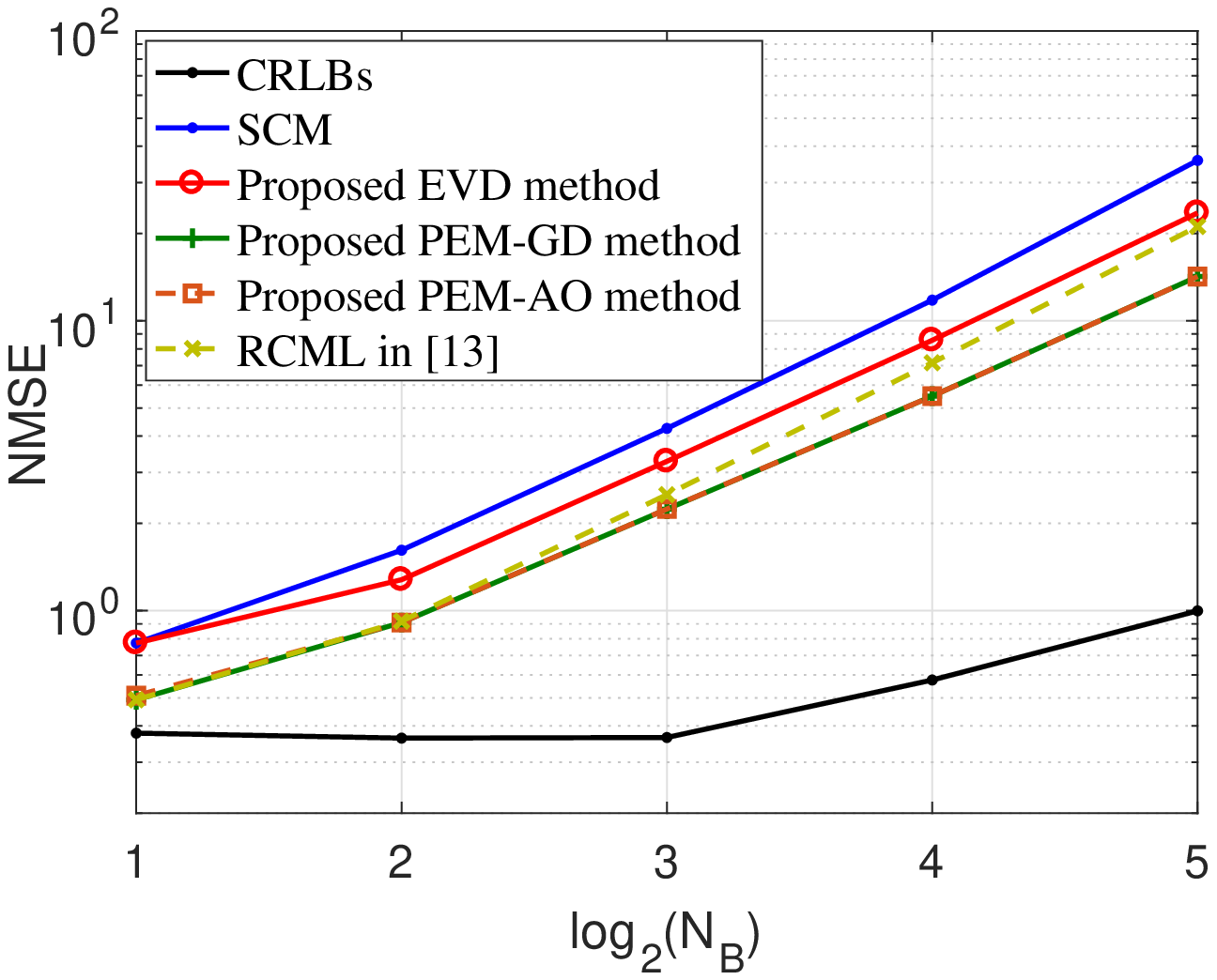}
		\caption{NMSE versus $N_B$}
	\end{minipage}
	\begin{minipage}[ht]{0.33\linewidth}
		\centering
		\includegraphics[width=2.56in]{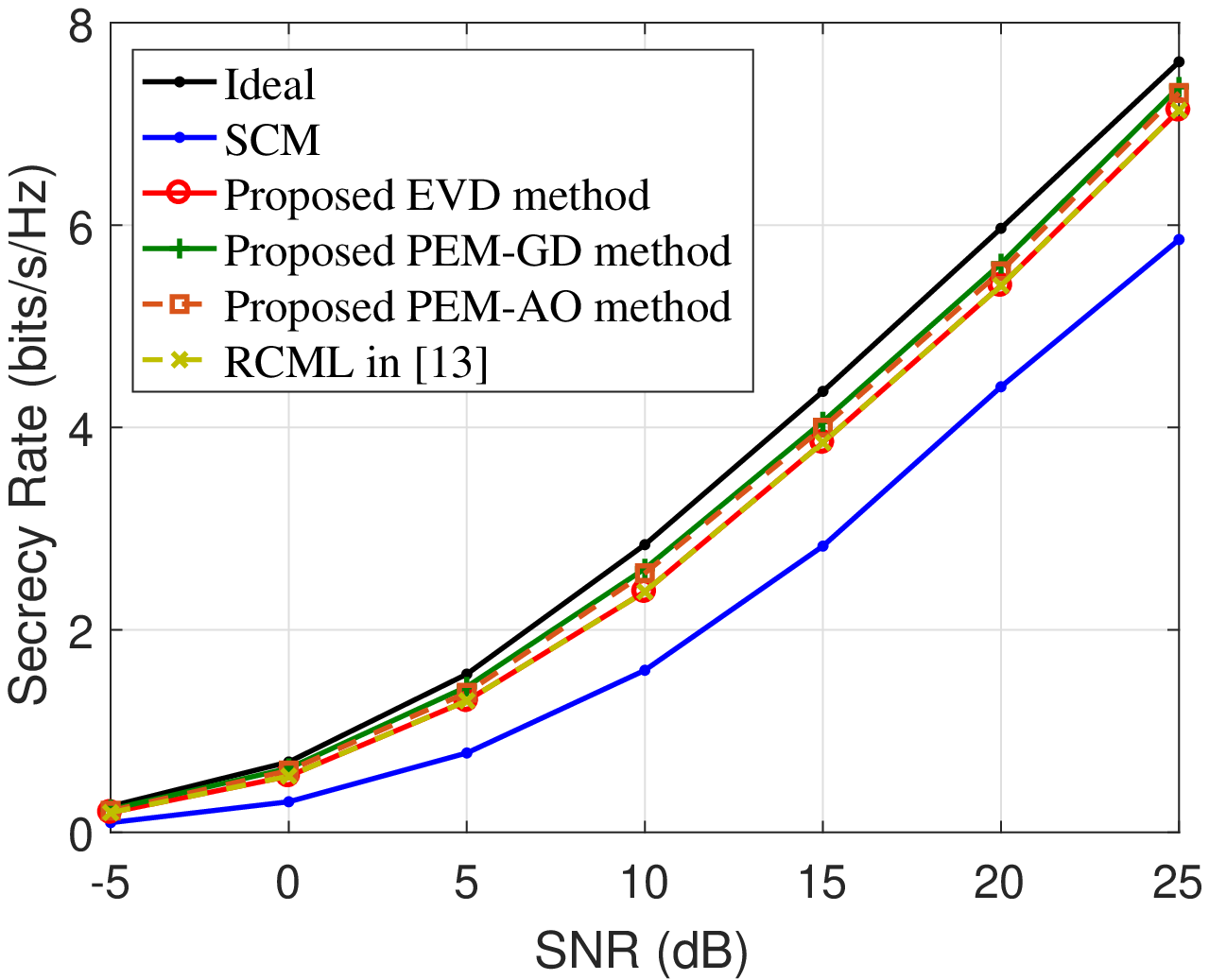}
		\caption{SR versus SNR}
	\end{minipage}
\vspace{-0.3cm}
\end{figure*}
\setcounter{figure}{1}
\begin{figure}[t]
\centering
\includegraphics[width=2.56in]{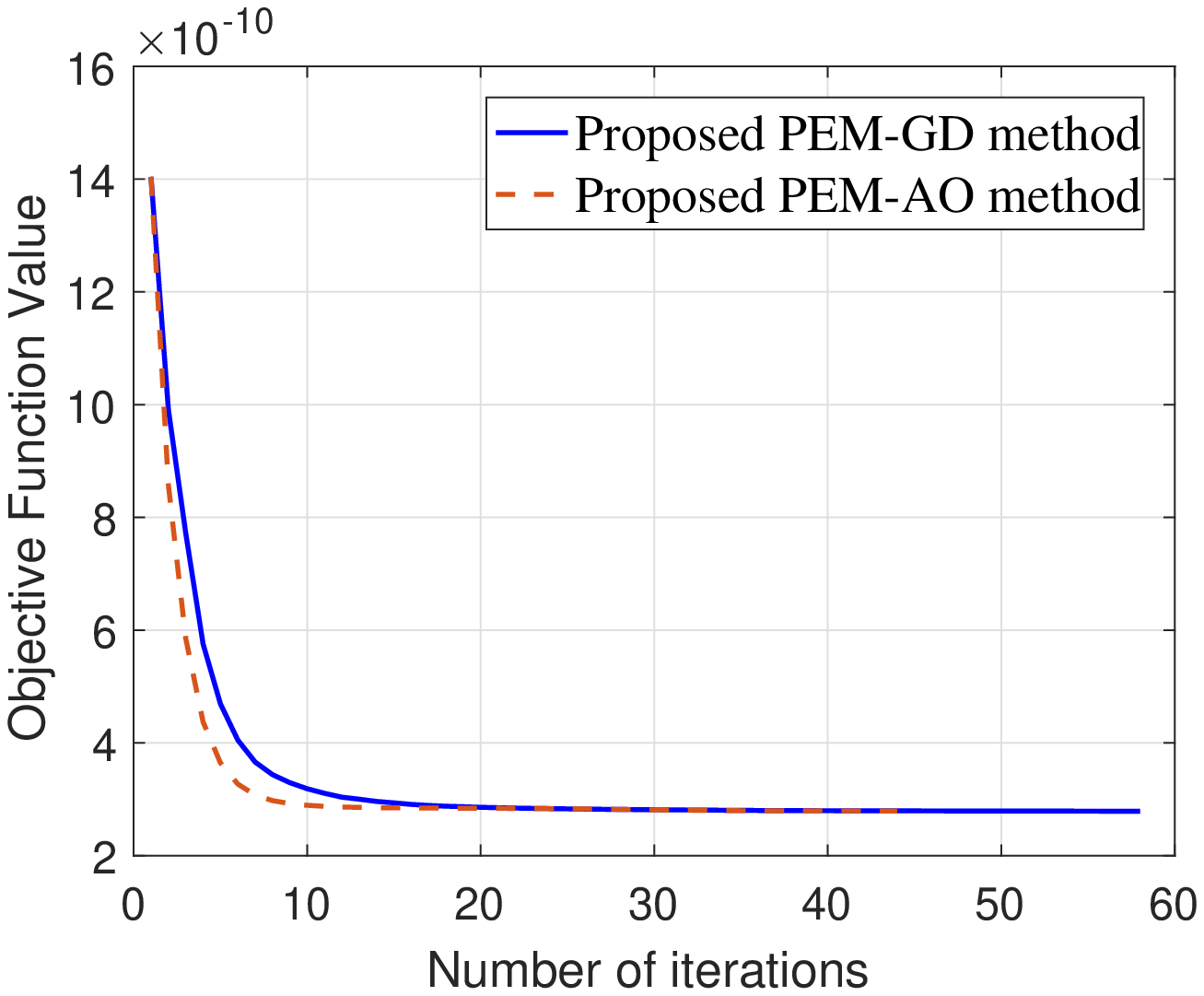}
\centering
\caption{Objective function value versus number of iterations}
\end{figure}

Fig.2 depicts the convergence of PEM-GD and PEM-AO. Obviously, the proposed PEM-GD and PEM-AO converge to approximate constant floor values, but the proposed PEM-AO has a faster convergence rate. Moreover, taking the calculation quantity of each iteration into account, we conclude that the proposed PEM-AO is of lower-complexity than PEM-GD.

Fig.3 plots the  NMSE versus JNR of three proposed methods with CRLBs, SCM and RCML in \cite{kang2014rank} as benchmarks. It is observed that in the low and medium regions of JNR, the proposed methods make better estimations than SCM while the RCML is better than EVD method but inferior to PEM-GD and PEM-AO. Additionally, all  methods tend to have the same performance as JNR increases to more than 10dB.

Fig.4 shows the  NMSE versus $N_B$. It is seen that as the dimension of the JCM matrix increases, with same amount of samples, the NMSEs of all methods will become worse. As well, their performance gaps are further widened. In other words,  the advantages of PEM-GD and PEM-AO over other methods will become more significant.

Fig.5 demonstrates the curves of SR versus SNR with different JCM in RBF. The figure shows that using JCM estimated by PEM-AO or PEM-GD achieves higher SR than other estimation methods, and EVD, RCML have approximate performance when applied to RBF.

\section{Conclusion}\label{conclusion}
In this paper, three methods: EVD, PEM-GD and PEM-AO have been proposed to estimate JCM before employing RBF methods to eliminate the active jamming from Mallory on Bob in an IRS-aided DM network. Simulation results showed that the three proposed methods perform better than SCM in the low and medium JNR regions in terms of NMSE and SR, while the proposed PEM-GD and PEM-AO outperform RCML and EVD. Among the three proposed methods, the proposed EVD is the lowest-complexity one and the complexity of PEM-AO is lower than PEM-GD. Among the three proposed methods, the proposed EVD is the lowest-complexity one and the complexity of PEM-AO is lower than PEM-GD.

\bibliographystyle{unsrt}
\bibliography{literature}

\begin{thebibliography}{10}
\providecommand{\url}[1]{#1}
\csname url@samestyle\endcsname
\providecommand{\newblock}{\relax}
\providecommand{\bibinfo}[2]{#2}
\providecommand{\BIBentrySTDinterwordspacing}{\spaceskip=0pt\relax}
\providecommand{\BIBentryALTinterwordstretchfactor}{4}
\providecommand{\BIBentryALTinterwordspacing}{\spaceskip=\fontdimen2\font plus
\BIBentryALTinterwordstretchfactor\fontdimen3\font minus
  \fontdimen4\font\relax}
\providecommand{\BIBforeignlanguage}[2]{{%
\expandafter\ifx\csname l@#1\endcsname\relax
\typeout{** WARNING: IEEEtran.bst: No hyphenation pattern has been}%
\typeout{** loaded for the language `#1'. Using the pattern for}%
\typeout{** the default language instead.}%
\else
\language=\csname l@#1\endcsname
\fi
#2}}
\providecommand{\BIBdecl}{\relax}
\BIBdecl

\bibitem{dm2009}
M.~P. Daly and J.~T. Bernhard, ``Directional modulation technique for phased
  arrays,'' \emph{IEEE Trans. Antennas Propag.}, vol.~57, no.~9, pp.
  2633--2640, 2009.

\bibitem{FDADM}
Q.~Cheng, V.~Fusco, J.~Zhu, S.~Wang, and F.~Wang, ``Wfrft-aided power-efficient
  multi-beam directional modulation schemes based on frequency diverse array,''
  \emph{IEEE Trans. Wireless Commun.}, vol.~18, no.~11, pp. 5211--5226, 2019.

\bibitem{robustDM}
J.~Hu, F.~Shu, and J.~Li, ``Robust synthesis method for secure directional
  modulation with imperfect direction angle,'' \emph{IEEE Commun. Lett.},
  vol.~20, no.~6, pp. 1084--1087, 2016.

\bibitem{Precise}
F.~Shu, X.~Wu, J.~Hu, J.~Li, R.~Chen, and J.~Wang, ``Secure and precise
  wireless transmission for random-subcarrier-selection-based directional
  modulation transmit antenna array,'' \emph{IEEE J. Sel. Areas Commun.},
  vol.~36, no.~4, pp. 890--904, 2018.

\bibitem{multibeam}
B.~Qiu, L.~Wang, J.~Xie, Z.~Zhang, Y.~Wang, and M.~Tao, ``Multi-beam index
  modulation with cooperative legitimate users schemes based on frequency
  diverse array,'' \emph{IEEE Trans. Veh. Technol.}, vol.~69, no.~10, pp.
  11\,028--11\,041, 2020.

\bibitem{wu2020}
Q.~Wu and R.~Zhang, ``Towards smart and reconfigurable environment: Intelligent
  reflecting surface aided wireless network,'' \emph{IEEE Commun. Mag.},
  vol.~58, no.~1, pp. 106--112, 2020.

\bibitem{DMlai}
L.~Lai, J.~Hu, Y.~Chen, H.~Zheng, and N.~Yang, ``Directional modulation-enabled
  secure transmission with intelligent reflecting surface,'' in \emph{IEEE Int.
  Conf. Inf. Commun. Signal Process., ICICSP}, 2020, pp. 450--453.

\bibitem{IRSDM}
F.~Shu, Y.~Teng, J.~Li, M.~Huang, W.~Shi, J.~Li, Y.~Wu, and J.~Wang, ``Enhanced
  secrecy rate maximization for directional modulation networks via irs,''
  \emph{IEEE Trans. Commun.}, pp. 1--1, 2021.

\bibitem{chengRIS}
X.~Cheng, Y.~Lin, W.~Shi, J.~Li, C.~Pan, F.~Shu, Y.~Wu, and J.~Wang, ``Joint
  optimization for ris-assisted wireless communications: From physical and
  electromagnetic perspectives,'' \emph{IEEE Trans. Commun.}, pp. 1--1, 2021.

\bibitem{yang2020intelligent}
H.~Yang, Z.~Xiong, J.~Zhao, D.~Niyato, Q.~Wu, H.~V. Poor, and M.~Tornatore,
  ``Intelligent reflecting surface assisted anti-jamming communications: A fast
  reinforcement learning approach,'' \emph{IEEE Trans. Wireless Commun.},
  vol.~20, no.~3, pp. 1963--1974, 2021.

\bibitem{teng2020low}
Y.~Teng, J.~Li, L.~Liu, G.~Xia, X.~Zhou, F.~Shu, J.~Wang, and X.~You,
  ``Low-complexity and high-performance receive beamforming for secure
  directional modulation networks against an eavesdropping-enabled full-duplex
  attacker,'' \emph{arXiv preprint arXiv:2012.03169}, 2020.

\bibitem{jiang2021efficient}
X.~Jiang, X.~Liu, R.~Chen, Y.~Wang, F.~Shu, and J.~Wang, ``Efficient receive
  beamformers for secure spatial modulation against a malicious full-duplex
  attacker with eavesdropping ability,'' \emph{IEEE Trans. Veh. Technol.},
  vol.~70, no.~2, pp. 1962--1966, 2021.

\bibitem{kang2014rank}
B.~Kang, V.~Monga, and M.~Rangaswamy, ``Rank-constrained maximum likelihood
  estimation of structured covariance matrices,'' \emph{IEEE Trans. Aerosp.
  Electron. Syst.}, vol.~50, no.~1, pp. 501--515, 2014.

\bibitem{boyd2004convex}
S.~Boyd, S.~P. Boyd, and L.~Vandenberghe, \emph{Convex optimization}.\hskip 1em
  plus 0.5em minus 0.4em\relax Cambridge university press, 2004.

\end{thebibliography}

\end{document}